\documentclass{article}
\usepackage{emulateapj,pstricks,apjfonts}

\def\chandra 	{{\em Chandra}\/}

\def\as		{$^{\prime\prime}$}
\def\kmsmpc	{~km$\;$s$^{-1}\,$Mpc$^{-1}$}
\def\cmsq	{~cm$^{-2}$}
\def\deg	{$^{\circ}$}
\def\kms	{~km$\;$s$^{-1}$}
\def\bi{\bfseries\itshape}

\begin{document}

\submitted{ApJ Letters in press}

\lefthead{NON-HYDROSTATIC GAS IN A1795}
\righthead{MARKEVITCH ET AL.}

\title{NON-HYDROSTATIC GAS IN THE CORE OF THE RELAXED GALAXY CLUSTER A1795}

\author{M. Markevitch$^{1}$, A. Vikhlinin$^{1}$, P. Mazzotta$^{2}$}

\affil{Harvard-Smithsonian Center for Astrophysics, 60 Garden St.,
Cambridge, MA 02138; maxim@head-cfa.harvard.edu}

\footnotetext[1]{Also IKI, Russian Academy of Sciences. 
~~~$^2$ ESA Fellow.}

\setcounter{footnote}{2}

\begin{abstract}

\chandra\ data on A1795 reveal a mild edge-shaped discontinuity in the gas
density and temperature in the southern sector of the cluster at
$r=60\;h^{-1}$ kpc. The gas inside the edge is 1.3--1.5 times denser and
cooler than outside, while the pressure is continuous, indicating that this
is a ``cold front'', the surface of contact between two moving gases. The
continuity of the pressure indicates that the current relative velocity of
the gases is near zero, making the edge appear to be in hydrostatic
equilibrium. However, a total mass profile derived from the data in this
sector under the equilibrium assumption, exhibits an unphysical jump by a
factor of 2, with the mass inside the edge being lower. We propose that the
cooler gas is ``sloshing'' in the cluster gravitational potential well and
is now near the point of maximum displacement, where it has zero velocity
but nonzero centripetal acceleration. The distribution of this
non-hydrostatic gas should reflect the reduced gravity force in the
accelerating reference frame, resulting in the apparent mass discontinuity.
Assuming that the gas outside the edge is hydrostatic, the acceleration of
the moving gas can be estimated from the mass jump, \mbox{$a\sim
800\;h\;\;{\rm km}\;{\rm s}^{-1}\;(10^8\; {\rm yr})^{-1}$}. The
gravitational potential energy of this gas that is available for dissipation
is about half of its current thermal energy. The length of the cool filament
extending from the cD galaxy (Fabian et al.) may give the amplitude of the
gas sloshing, $30-40\;h^{-1}$ kpc. Such gas bulk motion might be caused by a
disturbance of the central gravitational potential by past subcluster
infall.

\end{abstract}


\section{INTRODUCTION}

The hot gas that is filling clusters of galaxies delineates the
gravitational potential of the cluster dark matter. This enables a
derivation of the cosmologically important cluster masses from the X-ray
data (e.g., Bahcall \& Sarazin 1977; Sarazin 1988). Clusters form via
mergers that stir the gas (e.g., Roettiger, Loken, \& Burns 1997 and other
simulations), after which the cluster should be left alone for a few Gyr to
reach hydrostatic equilibrium for the mass measurement to work. Clusters
with regular X-ray morphology and the X-ray brightness strongly peaked on a
cD galaxy should have formed sufficiently long ago and are therefore
believed to be the safest candidates for X-ray mass measurements. For a
number of clusters, however, the X-ray-derived masses are systematically
lower than the independent gravitational lensing measurements (e.g.,
Miralda-Escud\'e \& Babul 1995 and later works). Both lensing and X-ray
measurements can suffer from systematic effects such as line of sight
substructure and inadequate X-ray modeling (e.g., Bartelmann
\& Steinmetz 1996; Allen, Ettori, \& Fabian 2001).

A1795 ($z=0.062$) is one of the most relaxed clusters in X-rays (e.g., Buote
\& Tsai 1996; Briel \& Henry 1996). It has had sufficient time to develop a
cool core (e.g., Edge, Stewart, \& Fabian 1992), although its precise nature
is debated (Tamura et al.\ 2001). Zooming in on the center with \chandra,
Fabian et al.\ (2001) discovered a 40\as\ filament of cool gas extending
from the cD galaxy, coincident with the optical H$\alpha$ filament (Cowie et
al.\ 1983). A total mass profile was derived from \chandra\ data by Ettori
et al.\ (2000).

Below we analyze the same \chandra\ data set and show that even in the
centers of such relaxed clusters as A1795, full hydrostatic equilibrium may
not have been achieved, resulting in considerable X-ray mass underestimates.
A similar finding was earlier reported for the otherwise relaxed cluster
RXJ1720+26 (Mazzotta et al.\ 2001a). We measure the bulk velocity and, for
the first time, acceleration of the gas near the cluster center by analyzing
a mild brightness edge, similar to those detected by \chandra\ in several
other clusters (Markevitch et al.\ 2000; Vikhlinin, Markevitch \& Murray
2001; Mazzotta et al.\ 2001ab). We use $H_0=100\,h$ \kmsmpc\ and $q_0=0.5$;
if not specified, confidence intervals are 68\% for one-parameter.

\begin{figure*}[tb]
\pspicture(0,14.3)(18.5,24)

\rput[tl]{0}(0.3,24.0){\epsfxsize=8.5cm \epsfclipon
\epsffile{img4.ps_dist}}

\rput[tl]{0}(9.8,24.0){\epsfxsize=8.5cm \epsfclipon
\epsffile{img.sm.ps_dist}}

\rput[bl]{0}(8.0,23.3){\large\bi a}
\rput[bl]{0}(17.5,23.3){\large\bi b}

\rput[tl]{0}(-0.1,15.8){
\begin{minipage}{18.5cm}
\small\parindent=3.5mm
{\sc Fig.}~1.---({\em a}) ACIS image in the 0.7--5 keV band. Pixels are
2\as, point sources are excluded. A mild brightness edge is seen
$60\;h^{-1}$ kpc south of the center.  ({\em b}) The same image smoothed by
a Gaussian with $\sigma=1''$ (brightness increases from gray to red to
green). The cool filament (Fabian et al.\ 2001) is seen in the center.
Dashed ellipse centered on the cD galaxy approximately delineates the edge.
Dashed lines show two 60\deg\ sectors used for extracting radial profiles.
\par
\end{minipage}
}
\endpspicture
\end{figure*}

\section{ANALYSIS}
\label{sec:analysis}

We have combined two archival ACIS-S observations of A1795 performed in
December 1999 and March 2000, limiting the analysis to the S3 chip. The
standard data cleaning resulted in the total useful exposure of 36 ks. The
background was modeled as in, e.g., Markevitch et al.\ (2000). The cluster
0.7--5 keV image is shown in Fig.\ 1. In the center, there is the gas
filament discovered by Fabian et al.\ (2001). South of the center, outside
the filament region, one can see a subtle edge feature that is the subject
of this work.

The radial brightness profile, extracted in the southern 60\deg\ sector
centered on the cD galaxy (see Fig.\ 1{\em b}) is shown in Fig.\ 2{\em a}.
As other profiles below, it is derived in narrow elliptical segments
parallel to the edge (the ellipse in Fig.\ 1{\em b}), and plotted as a
function of the average emission-weighted radius in each segment. In the
narrow sectors of interest, these segments are reasonably close to circular,
so below we assume spherical symmetry for simplicity of our qualitative
analysis. The brightness profile has the characteristic shape of a projected
spherical gas density discontinuity. In the radial range $40''-140''$
(within a factor of 2 of the edge radius), the profile is fit very well by a
gas density model consisting of two power laws ($1.10\pm 0.06$ and $1.42\pm
0.03$) and a jump by a factor of $1.28\pm 0.03$ (see Fig.\ 2{\em c}).  The
observed temperature differences (described below) have a negligible effect
on the density model. Closer to the center lies the filament region, so we
do not include smaller radii in the analysis and show them only for
illustration.  At larger radii, the density profile is known to steepen
(Vikhlinin, Forman, \& Jones 1998) but this is unimportant for our fits.

For spectral modeling, we averaged the position-dependent telescope
effective area, the detector nonuniformity and spectral response over a
given region with the cluster brightness as weight. Response matrices
released on 2001 August 8 were used. The temperature profile in the southern
sector, derived using Galactic absorption and the Kaastra (1992) plasma
model, is shown in Fig.\ 2{\em b}. For the bins inside the edge, we have
subtracted the projected contribution from the outer, hotter gas, assuming a
6.7 keV temperature (the average for the 3 outer bins) and the best-fit
density model. The resulting profile is consistent with two constants inside
and outside the edge; thus, any more complex deprojection is not
warranted. The outside temperature is in agreement with the 6--7 keV average
observed at larger radii (e.g., Markevitch et al.\ 1998; Arnaud et al.\
2001).

Since our analysis includes regions with short cooling time, we checked if
any multicomponent gas could bias our measurements. We first tried to fit
our spectra with a simple two-temperature or standard cooling flow model,
which did not improve the fits. On the other hand, the spectrum of the
central region ($r<30''$) exhibits lines of ionized Si and S at 1.9 keV and
2.4 keV (rest frame) that can be fit with a heavily-absorbed ($N_H>
10^{22}$\cmsq), cool ($T\sim 0.5$ keV) thermal component, in addition to the
main 3--4 keV gas.  The origin of this emission is beyond the scope of this
paper, but we have looked for a similar contaminant in the regions of our
interest. For this, only the second observation taken at the ACIS
temperature of $-120^\circ$C was used, for which the current response
matrices have adequate accuracy (the $-110^\circ$C matrices are sufficient
to fit the continuum shape but not the lines). In the combined 3 spatial
bins outside the edge (solid crosses in Fig.~2{\em b}), such a component is
not required. For the inner 3 bins, the fit is marginally improved ($\Delta
\chi^2=8$ for 3 additional d.o.f.); however, the best-fit temperature of the
main gas does not change and the fit does not allow a significant filling
factor for these cool clouds. Also, the cool component gives only 2\% of the
flux in the 0.7--5 keV band and cannot bias our density profile for the main
gas. Thus, the derived quantities for the diffuse gas in our regions of
interest should not be affected.

\begin{figure*}[tb]
\pspicture(0,-0.5)(18.5,15.3)

\rput[tl]{0}(0.2,15.9){\epsfxsize=9.5cm \epsfclipon
\epsffile{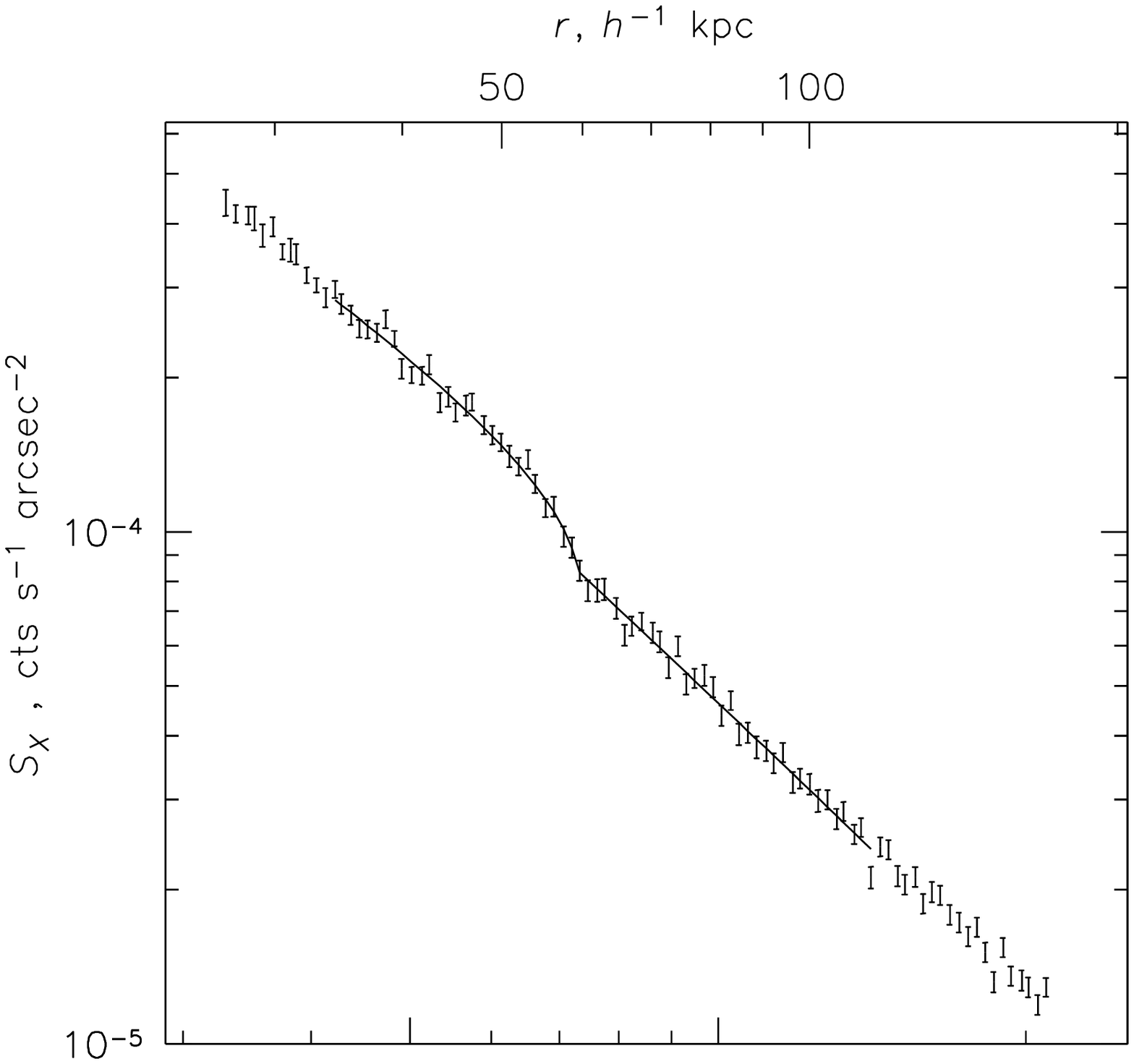}}

\rput[tl]{0}(0.2,10.5){\epsfxsize=9.5cm \epsfclipon
\epsffile{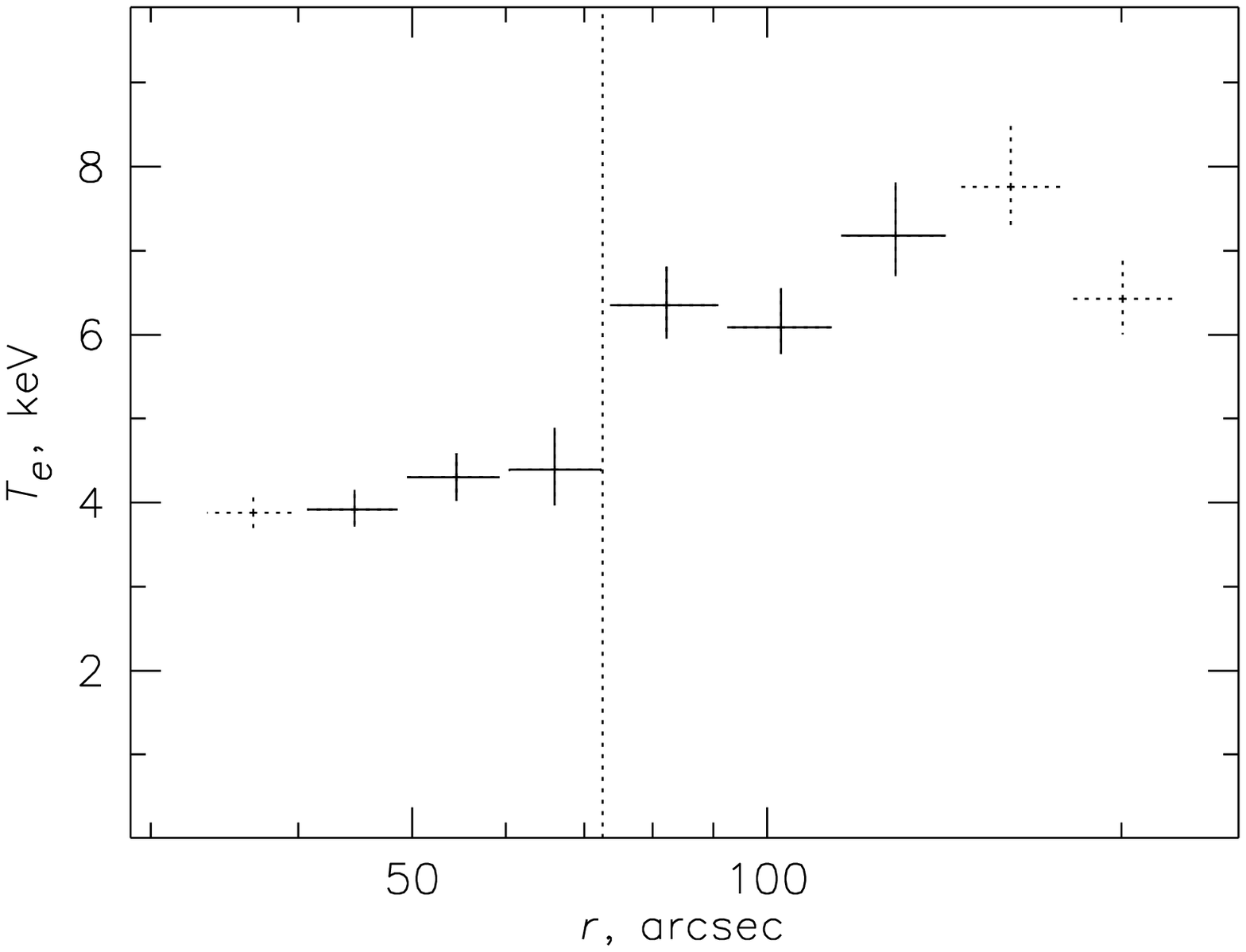}}

\rput[tl]{0}(9.4,16.18){\epsfxsize=9.5cm \epsfclipon
\epsffile{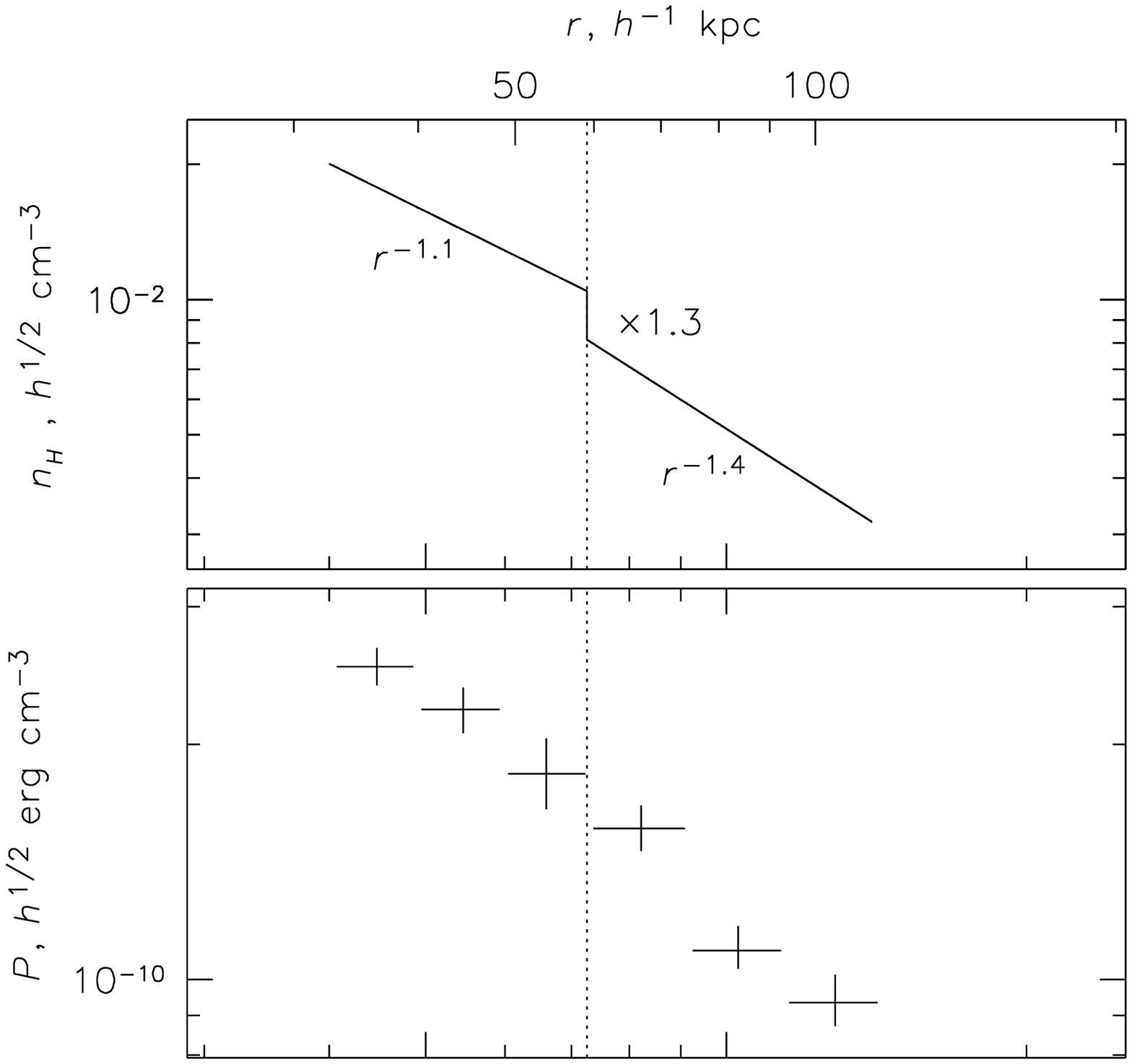}}

\rput[tl]{0}(9.4,10.5){\epsfxsize=9.5cm \epsfclipon
\epsffile{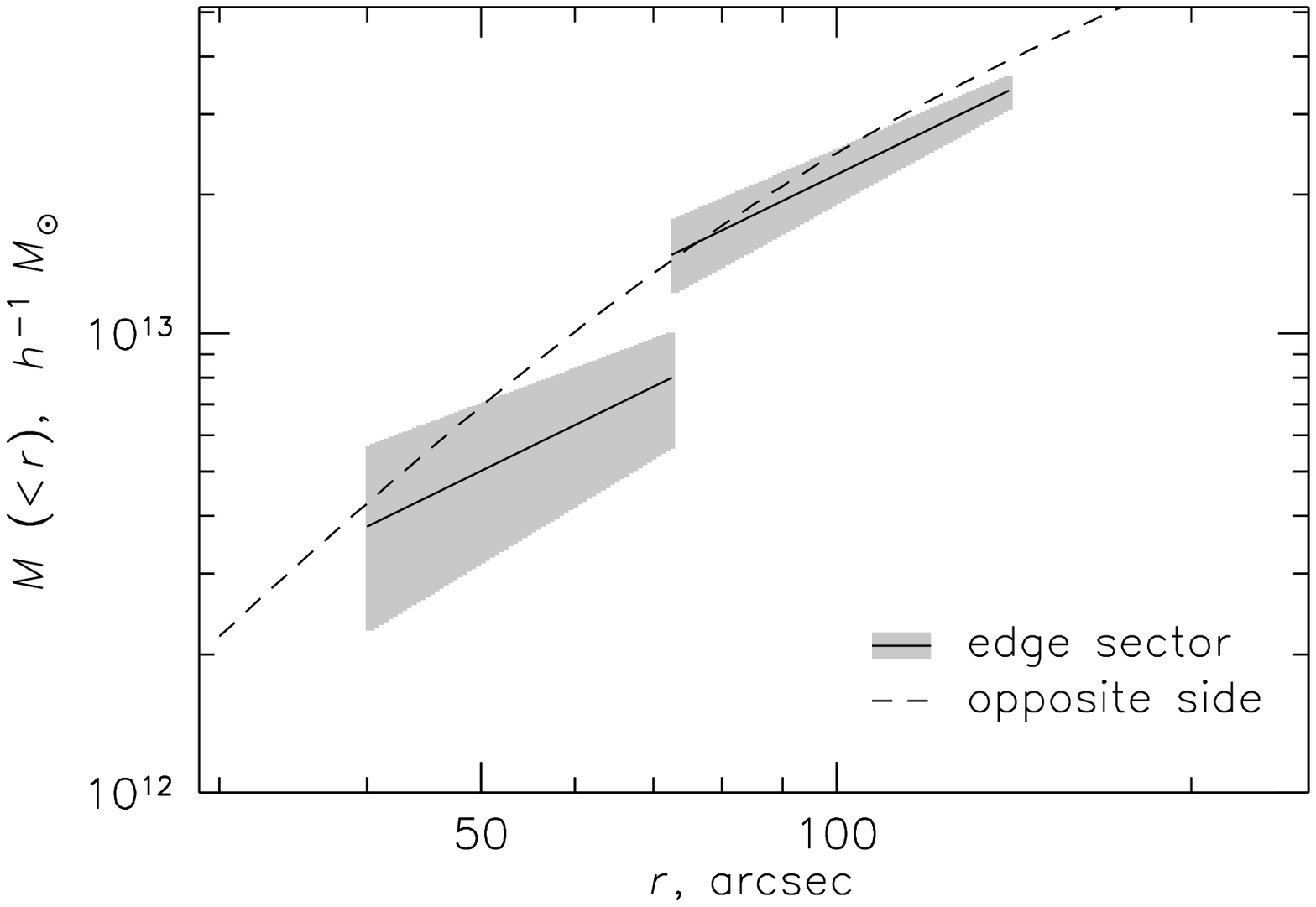}}

\rput[bl]{0}( 8.1,14.0){\large\bi a}
\rput[bl]{0}( 8.1, 2.8){\large\bi b}
\rput[bl]{0}(17.3,14.0){\large\bi c}
\rput[bl]{0}(17.3,10.5){\large\bi d}
\rput[bl]{0}(17.3, 6.8){\large\bi e}

\rput[tl]{0}(-0.1,1.4){
\begin{minipage}{18.5cm}
\small\parindent=3.5mm
{\sc Fig.}~2.---Profiles from the southern 60\deg\ sector containing the
edge.  Errors are 68\%. Vertical dotted lines show the edge position. ({\em
a}) Surface brightness in the 0.7--5 keV band. The line is a fit in the
$r=40''-140''$ range; the corresponding density model is shown in panel
({\em c}). ({\em b}) Temperature profile with approximate deprojection (see
text). Solid crosses show values used for derivation of the pressure and
total mass across the edge. ({\em d}) Pressure profile.  ({\em e}) Total
mass within a given radius (confidence bands are shown in gray).  For
comparison, a fit in the sector opposite to the edge (see Figs. 1{\em b} and
3) is shown by dashed line; they must show the same mass if the gas is in
hydrostatic equilibrium.
\par
\end{minipage}
}
\endpspicture
\end{figure*}

\section{DISCUSSION}

The temperature profile reveals that the edge is a ``cold front'', or a
contact discontinuity between two moving gases, similar to (although not
quite as prominent as) those discovered by \chandra\ in several other
clusters. From the edge geometry, the denser gas should be moving away from
the center (or the outer gas flowing onto the center).  Multiplying the gas
density by its temperature, we obtain a profile of thermal pressure across
the edge (Fig.\ 2{\em d}). It is continuous; if we fit the inner 3 and outer
3 temperature values (solid crosses in Fig.\ 2{\em b}) by power law
functions of radius, the pressure jump is $1.0\pm0.2$ (a 90\% error
including the scatter of the power law slopes). This leaves little or no
room for a relative gas motion, since the ram pressure from such a motion
would cause the inner thermal pressure to be higher compared to that on the
outside (beyond the small stagnation region), as is indeed seen in A3667
(Vikhlinin et al.\ 2001) The above upper limit on the pressure jump
corresponds to a 90\% upper limit on the Mach number of 0.5 (see, e.g.,
Fig.\ 6 in Vikhlinin et al.\ 2001) The limit is probably even stronger,
because by fitting the temperature profiles with power laws in narrow radial
intervals, we may allow too much unphysical freedom. If instead one assumes
constant temperatures inside and outside, the Mach number should be very
close to 0.

Thus the inner and outer gases appear very nearly at rest and in pressure
equilibrium. Therefore, one expects them to be in hydrostatic equilibrium in
the cluster gravitational potential. Under this assumption and that of
spherical symmetry, we can derive the cluster total mass profile in this
radial range using our power-law fits to the temperature and density
profiles in the immediate vicinity of the edge. This mass profile is shown
in Fig.\ 2{\em e} (the error bands include uncertainties of the local
temperature gradients; uncertainties on the density gradients are negligible
in comparison). The profile reveals an unphysical discontinuity by a factor
of 2, with a formal significance of 99\%. For comparison, we analyzed the
northern sector of the cluster opposite to the edge. A brightness profile
there shows no features and is fit well by the customary $\beta$-model
(Fig.\ 3{\em a}).  The temperature is nearly constant, but for exactness, we
have modeled its slow radial rise by a smooth function. Its
emission-weighted projection is overlaid on the data in Fig.\ 3{\em b}. The
total mass profile derived using these fits is shown by dashed line in Fig.\
2{\em e}.  Outside the edge radius, the masses derived from the two opposite
sectors agree, as they must. This indicates that the gas immediately outside
the edge is indeed near hydrostatic equilibrium, while the gas inside the
edge is not.

Similar mass discontinuities at the cold fronts were observed in A3667
(Vikhlinin et al.\ 2001) and RXJ1720+26 (Mazzotta et al.\ 2001a). However,
in the former, the cold front is a boundary of a subcluster flying with the
sonic velocity where one does not expect hydrostatic equilibrium. The latter
cluster is apparently relaxed on large scales so the situation is similar to
A1795, but there, too, a high velocity of the front could not be excluded.
Mazzotta et al.\ (2001a) proposed that the core of RXJ1720+26 is moving
relative to the rest of the cluster as a result of the peculiar history of
that cluster's collapse. However, such an explanation for A1795 (as well as
several other generally relaxed clusters with similar features, M.
Markevitch et al.\ in preparation) may require too much of a coincidence.

We propose instead that the cool gas just inside the edge is ``sloshing''
north-south in the central potential well (which may be static at present)
and is now observed near the point of maximum displacement, where it has
zero velocity but nonzero centripetal acceleration. Such a gas would not be
in hydrostatic equilibrium with the potential --- instead, its distribution
would reflect the reduced gravity force in the accelerating reference frame,
resulting in the apparent mass underestimate. Assuming that the gas outside
the edge is hydrostatic, the acceleration of the moving gas can be estimated
from the mass jump $\Delta M$ as $a\sim G \Delta M r^{-2} \approx 3\times
10^{-8}\,h$ cm~s$^{-2}$ or $800\;h\;\;{\rm km}\;{\rm s}^{-1}\;(10^8\;{\rm
yr})^{-1}$, where $r$ is the radius of the edge.

It is interesting to estimate the gravitational potential energy of this
gas. Approximately, $E\sim G\,\Delta M r^{-1} \approx 5\times
10^{15}$~erg~g$^{-1}$. For a 4 keV gas, this is about half of its thermal
energy. As the gas accelerates, this energy becomes available for
dissipation. Thus, such sloshing can produce significant heating in the
cluster center that may offset some radiative cooling.

Unlike A3667 (the site of the most prominent cold front) where the cool gas
is retained by a merging subcluster and travels together with its distinct
dark matter halo (A. Vikhlinin et al.\ in preparation), here the moving gas
is probably not attached to any underlying moving dark matter body. Indeed,
simulations of cluster mergers (e.g., Roettiger et al.\ 1997) predict that
late into a merger, the gas decouples from dark matter and sloshes in the
changing gravitational potential for a significant time before coming to a
hydrostatic equilibrium.

\noindent
\pspicture(0,0.4)(9,14.1)

\rput[tl]{0}(-0.2,15.6){\epsfxsize=9.cm \epsfclipon
\epsffile{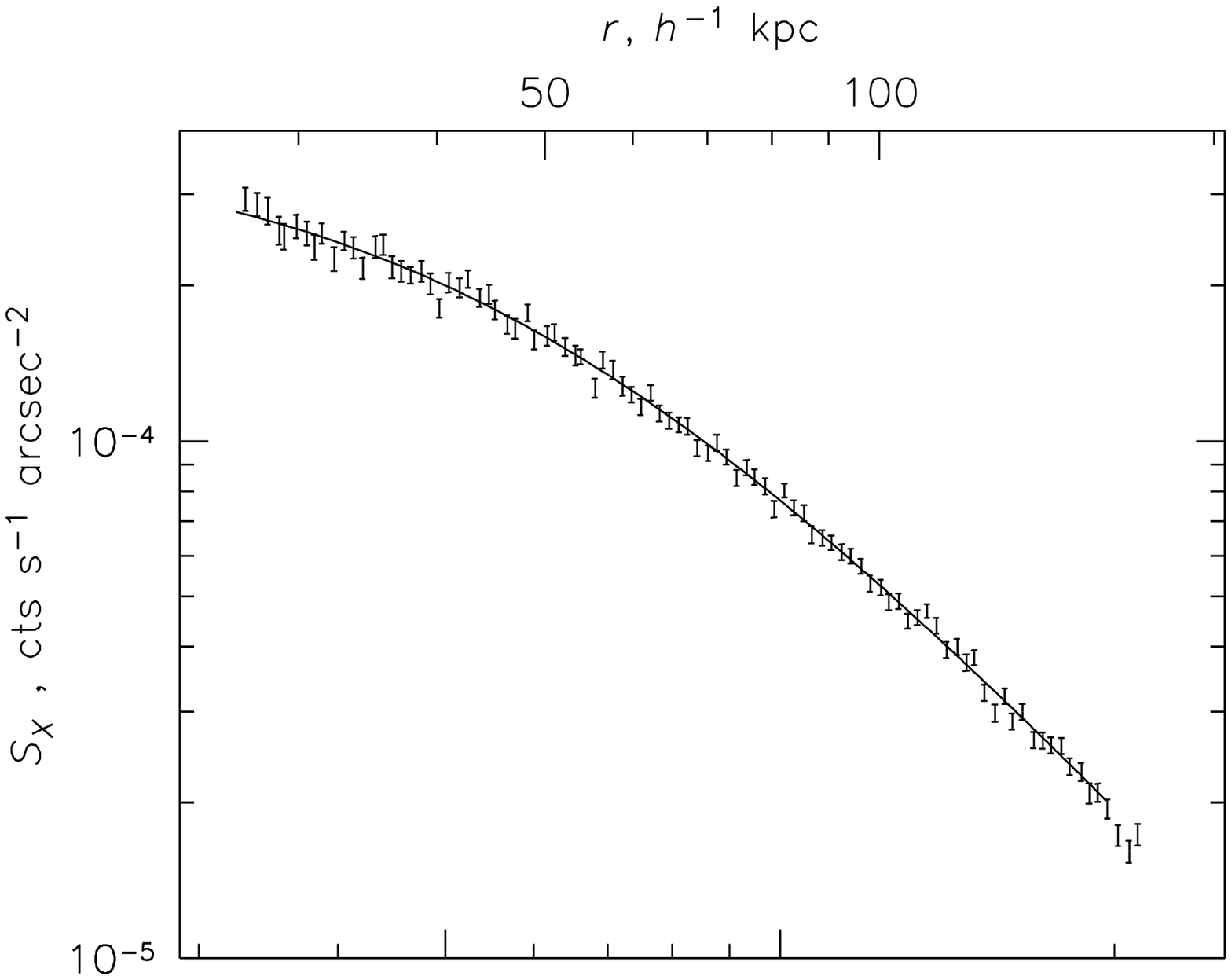}}

\rput[tl]{0}(-0.2,10.8){\epsfxsize=9.cm \epsfclipon
\epsffile{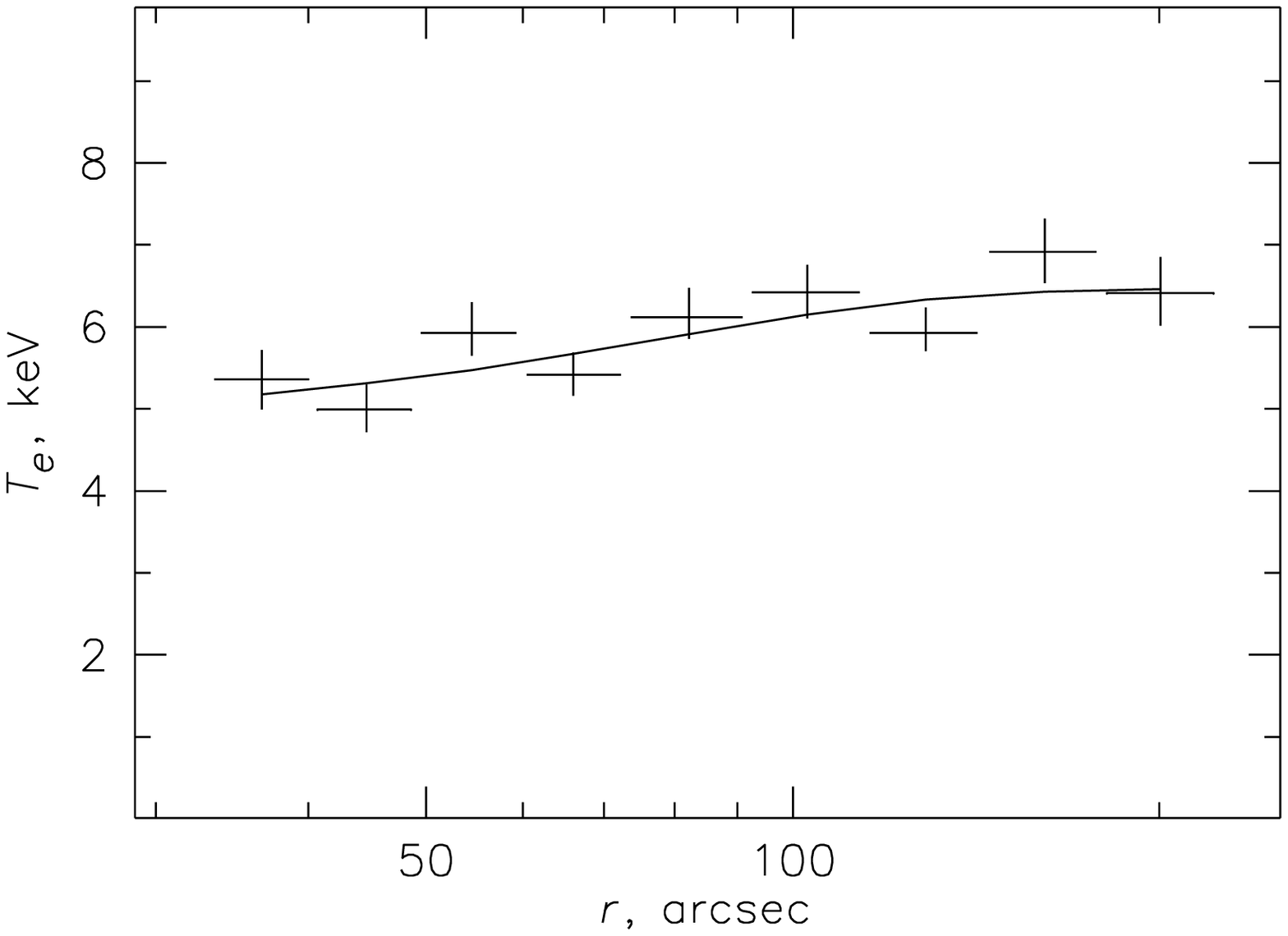}}

\rput[bl]{0}( 7.25,12.6){\large\bi a}
\rput[bl]{0}( 7.25, 3.5){\large\bi b}

\rput[tl]{0}(0,2.2){
\begin{minipage}{8.75cm}
\small\parindent=3.5mm
{\sc Fig.}~3.---Profiles from the northern 60\deg\ sector opposite to the
edge.  ({\em a}) Surface brightness. The line shows a fit by a
$\beta$-model.  ({\em b}) Projected temperature profile. The line shows a
best-fit model used to estimate the total mass (Fig.~2{\em e}). Errors are
68\%.
\par
\end{minipage}
}
\endpspicture

However, a recent major merger in A1795 is unlikely because of the
exceptional regularity of its X-ray image. Also, the gas temperature
decreases toward the center, indicating that either the gas has had time to
cool ($t_{\rm cool}\approx 10^{10}$ yr at the edge radius) or the cool gas
from subcluster infall has had time to settle down, both processes requiring
the absence of significant disturbance. Perhaps a localized gas motion in
the core might be caused by a disturbance of the central gravitational
potential by an infalling small subcluster. Such a disturbance would supply
kinetic energy to the low-entropy gas accumulated over time in the cluster
center, which is more responsive to small potential perturbations than the
hotter outer gas. Alternatively, in clusters where the central AGN produces
large bubbles of hot plasma (e.g., Churazov et al.\ 2000; McNamara et al.\
2000), such activity could supply kinetic energy to the surrounding cool
gas. There are no obvious large bubbles in A1795.

We note that the central cool filament, extending southward from the cD
galaxy (Fig.\ 1{\em b}), is aligned with the apparent direction of the most
recent gas movement. This suggests that the cD galaxy in A1795 does not move
with the gas but probably stays with the gravitational potential (to within
its observed small 150\kms\ velocity relative to the cluster average;
Oegerle \& Hill 1994). The filament is probably a cooling wake arising as
the cD galaxy is moving relative to the gas, as proposed by Fabian et al.\
(2001), only their relative motion is not due to the cD galaxy's
oscillations in the potential but rather to the sloshing of the gas. The
filament length, $30-40\;h^{-1}$ kpc, should then give the amplitude of the
gas oscillations in the central potential well.

The observed mass discontinuity may have significant effect on X-ray
measurements of the total mass at the cluster centers. It is at these radii
where strong lensing often implies 2--3 times higher values (e.g.,
Miralda-Escud\'e \& Babul 1995 and later works), although more accurate
X-ray modeling was shown to reduce the discrepancy (e.g., Allen et al.\
2001). Given the filament, the edge and the overall image geometry, it
appears likely that most of the gas at small radii in A1795 participates in
the gas bulk flow and is therefore nonhydrostatic, even though we can see
the effect on the mass profile only in the southern sector where sloshing
produces a brightness edge. \chandra\ data on a number of other relaxed
clusters reveal similar fronts in the cluster central regions (M. Markevitch
2001, in preparation) so such nonequilibrium may be common. Note that the
volume inside the edge radius contains only $\sim 1$\% of the gas within the
virial radius, so our finding cannot be generalized to the whole cluster.

In summary, we have studied a mild gas density edge near the center of the
otherwise relaxed cluster A1795. The gas on two sides of the edge appears at
rest; however, the inner gas has significant centripetal acceleration,
indicating that the gas is sloshing in the cluster gravitational potential.
If one assumes hydrostatic equilibrium, one obtains an underestimate of the
total mass near the center.

\acknowledgements

We thank Harvey Tananbaum for useful discussions. 
Support for this study was provided by NASA contract NAS8-39073, NASA grant
NAG5-9217 and ESA fellowship.

\end{document}